\documentclass[twocolumn,showpacs,amsmath,amssymb,aps,pra,floatfix,nofootinbib,10pt]{revtex4-2}
\usepackage[latin1]{inputenc}
\usepackage[english]{babel}
\usepackage[T1]{fontenc}
\usepackage{mathrsfs}
\usepackage{hyperref}
\usepackage{multirow}
\usepackage{bm,epsfig}
\usepackage{graphicx}
\usepackage{subeqnarray}
\usepackage{bbold}
\usepackage{upgreek}
\hypersetup{colorlinks=true, urlcolor=blue, citecolor=blue, filecolor=blue, linkcolor=blue}
\usepackage{dcolumn}
\usepackage{amsmath}
\usepackage{color}
\usepackage{siunitx}
\usepackage{physics}

\def\fig#1{Fig.\,\ref{#1}}

\def\beq#1{\begin{equation}\label{#1}}
\def\eeq{\end{equation}}

\begin{document}

\title{
Generation of Attosecond Pulses with Controllable Carrier-Envelope Phase \\ via High-order Frequency Mixing
}

\author{
V.~A.~Birulia$^1$, M.~A.~Khokhlova$^{2,3,\ast}$, V.~V.~Strelkov$^{1}$
}
\affiliation{
\mbox{$^{1}$Prokhorov General Physics Institute of the Russian Academy of Sciences, Vavilova street 38, Moscow 119991, Russia} \\
\mbox{$^{2}$Max Born Institute for Nonlinear Optics and Short Pulse Spectroscopy, Max-Born-Stra{\ss}e 2A, Berlin 12489, Germany} \\
\mbox{$^{3}$King's College London, Strand, London WC2R 2LS, UK} \\
$^{\ast}$margarita.khokhlova@kcl.ac.uk
}

\begin{abstract}
Advancing table-top attosecond sources in brightness and pulse duration is of immense interest and importance for an expanding sphere of applications. 
Recent theoretical studies [New J.\ Phys., \textbf{22} 093030 (2020)] have found that high-order frequency mixing (HFM) in a two-color laser field can be much more efficient than high-order harmonic generation (HHG). 
Here we study the attosecond properties of the coherent XUV generated via HFM analytically and numerically, focusing on the practically important case when one of the fields has much lower frequency and much lower intensity than the other one. 
We derive simple analytical equations describing intensities and phase locking of the HFM spectral components. 
We show that the duration of attosecond pulses generated via HFM, while being very similar to that obtained via HHG in the plateau, is shortened for the  cut-off region. 
Moreover, our study demonstrates that the carrier-envelope phase of the attopulses produced via HFM, in contrast to HHG, can be easily controlled by the phases of the generating fields.

\end{abstract}

\maketitle
\noindent

\section{Introduction}
The introduction of laser technologies ignited an explosion in the study of light and its interaction with matter. One of the areas from this realm is attosecond physics~\cite{Corkum2007, Krausz2009, Villeneuve2018, Ryabikin2022}. In turn, the rapid expansion of attophysics and the conquest of progressively ultrafast processes has created a hunger for attosecond sources and their active development, resulting in steady progress~\cite{Johnson2018Aug, Gaumnitz2017Oct, Li2020Jun, Ye_2020}. However, there is still a high demand for further efforts dictated by the needs of a growing sphere of applications. 

Currently available table-top attosecond sources are based on the generation of high-order harmonics of intense laser pulses during their interaction with a gaseous medium. 
A process similar to high-order harmonic generation (HHG) occurs when two fields~--- at least one of which is intense and therefore causes photoionization of the gas~--- generate high-order mixed-frequency components. This process is called high-order frequency mixing (HFM)~\cite{Eichmann1995, Kapteyn_2007, Worner_non-collinear, Oguchi_PRA, Strelkov_np,Ganeev2016,Ellis2017,Tran2019,Harkema2019,Chappuis2019,Jiang2021}.

The efficiency of the macroscopic HHG response is substantially limited by the phase matching of the process. Namely, the process is indissolubly connected to the photoionization of the medium, and the change in the refractive index due to the ionization leads to weaker phase matching~\cite{Constant}. This limitation can be significantly softened for the HFM process under a proper choice of frequencies of the generating fields~\cite{Shkolnikov, Chichkov_plasma, Platonenko_non-collinear, Chichkov_2000, PhysRevLett.112.143902}, resulting in much longer propagation distances of the phase-matched generation (including the case when second field has much weaker intensity and much lower frequency than the fundamental) and thus to higher efficiency of the HFM~\cite{KhokhlovaStrelkov, Hort2021}. 
This advantage defines the perspective of using the HFM process to design highly-effective attosecond pulse sources. The scope of the current paper is theoretical investigation of these perspectives. 

In this paper we study theoretically the microscopic aspects of attosecond pulse generation via the HFM process. Our analytical approach is based on the strong-field approximation (SFA)~\cite{Lewenstein} extended to include the second weak field perturbatively~\cite{Gaarde1996}. Assuming that this field can be considered as a quasi-static one, we derive simple analytical equations for the amplitudes and phases of the HFM components. The analytical SFA results are compared with ones of the numerically-integrated SFA and also with results obtained via numerical simulation of the three-dimensional time-dependent Schr\"odinger equation (3D TDSE). 

An intrinsic feature of the HHG atomic response is its frequency modulation, or the ``attochirp''~\cite{Lewenstein1995, Antoine1996, Mairesse2003, Kohler2011}. It defines the lower limit for the duration of the attosecond pulse in the plateau region~\cite{Strelkov2016}. Here we study the attochirp of the pulses obtained via HFM both for the plateau and the cut-off regions. 

The phase of the carrier with respect to the pulse envelope, or the carrier-envelope phase (CEP), is a key feature of the few-cycle pulses. The ability to stabilize it via $f$-$2f$ interferometry~\cite{Paschotta} and to control it leads to numerous new perspectives in studies of the interaction of intense femtosecond pulses with matter. However, the CEP of the attosecond pulses obtained via HHG cannot be controlled easily. In this paper we show that HFM allows a straightforward way to control the CEP of the attosecond pulse via tuning the phases of the generating fields.

\section{Analytical Theory}
In this section we study analytically the microscopic response of a model atom to a two-color linearly-polarized field in the framework of the strong-field approximation~\cite{Lewenstein}. The two-color field consists of an intense laser field and a weaker low-frequency field which we assume for the moment to be static, so the total field is written as
\beq{E}
E(t)=\mathcal {E}_0 \cos(\omega_0 t)+\mathcal{E}_1 \, , 
\eeq
where the amplitudes of the strong field $\mathcal{E}_0$ and of the weak field $\mathcal{E}_1$ satisfy the condition
\beq{condition_amp}
\mathcal{E}_1 \ll \mathcal {E}_0 \, .
\eeq
We write the time-dependent dipole moment derived as the integral~(13) in~\cite{Lewenstein} in the form
\beq{x}
x(t)= \int_0^\infty d \tau f (t, \tau) \exp( -i S_\mathrm{st}(t, \tau)) + \mathrm{c.c.} \, , 
\eeq
where $S_\mathrm{st}(t, \tau)$ is the quasi-classical action, and $f (t, \tau)$ denotes the remaining part of the integrand excluding the exponent $\exp( -i S_\mathrm{st})$. 

The quasi-classical action is given as
\beq{S_st_init}
S_\mathrm{st} (t, \tau)=\frac{1}{2} \int_{t-\tau}^t dt'' (p_\mathrm{st}-A(t''))^2 \, ,
\eeq
where 
\beq{A}
A(t)=-\frac{\mathcal{E}_0}{\omega_0}\sin{(\omega_0 t)} -\mathcal{E}_1 t
\eeq
is the vector potential of the field~\eqref{E}
and $p_\mathrm{st}$ is the stationary value of the momentum, which allows the electron trajectory starting near the origin at the time instant $t-\tau$ to return to the same position at the time instant $t$. This stationary value of the momentum in the field~\eqref{E} is written 
\beq{p_st}
p_\mathrm{st}=p_\mathrm{st}^{(0)}+ (\tau / 2 - t) \mathcal{E}_1 \, ,
\eeq
where $p_\mathrm{st}^{(0)}$ is the stationary value of the momentum in the absence of the second field,%
\footnote{%
Here and below we use the upper index~$^{(0)}$ to denote values in the absence of the second field.
}\ %
given by Eq.~(14) in Ref.~\cite{Lewenstein}, which we rewrite in atomic units%
\footnote{%
Here we use standard atomic units in contrast to Ref.~\cite{Lewenstein}, where in addition to the use of atomic units all energies are expressed in terms of the laser photon energy. 
}\ %
as
\beq{p_st0}
p_\mathrm{st}^{(0)}=\frac{\mathcal{E}_0}{\omega_0^2 \tau} \left[ \cos{(\omega_0 t)} -\cos{(\omega_0 t- \omega_0 \tau)} \right] \, .
\eeq

Substituting Eq.~(\ref{p_st}) in Eq.~(\ref{S_st_init}), we derive the action~as
\beq{S_st}
S_\mathrm{st}=S_\mathrm{st}^{(0)}- \frac{\mathcal{E}_1}{\mathcal{E}_0} \, \frac{ 2 U_p}{\omega_0} D (t,\tau) + \frac{\mathcal{E}_1^2 \tau^3}{24} \, ,
\eeq
where 
\beq{S_st_0}
\begin{split}
S_\mathrm{st}^{(0)} = & U_p \tau 
- 
\frac{2 U_p}{\omega_0^2 \tau}
\left(
1-\cos{(\omega_0 \tau)} 
\right) - \\ &
\frac{U_p}{\omega_0} 
\cos{
\left(
\omega_0 (2t-\tau)
\right)
}
\left[
\sin{(\omega \tau)} - 
\frac{4\sin^2{(\omega_0 \tau / 2)}}{\omega_0 \tau} 
\right]
\end{split}
\eeq
is the action found in~\cite{Lewenstein} for a single-color field,
\beq{D}
\begin{split}
D(t,\tau)= \ &2[\sin (\omega_0 (t- \tau)) - \sin(\omega_0 t)]+ \\
&\omega_0 \tau [\cos (\omega_0 (t- \tau)) + \cos(\omega_0 t)] \, 
\end{split}
\eeq
and $U_p=\mathcal{E}_0^2/{4\omega_0^2}$ is the ponderomotive energy in the strong field $\mathcal{E}_0$. Expression~(\ref{S_st}) presents the action in the field~\eqref{E} with vector potential~\eqref{A} as a quadratic polynomial in $\mathcal{E}_1$.

After expanding the exponent $\exp( -i S_\mathrm{st})$ up to the term proportional to $\mathcal{E}_1^2$ within the new action~(\ref{S_st}), we obtain
\beq{exp_S_st}
\begin{split}
&\exp(-i S_\mathrm{st})=\exp(-i S_\mathrm{st}^{(0)}) \times \\
&\left[ 
1
+ i \frac{\mathcal{E}_1}{\mathcal{E}_0} \frac{2U_p}{\omega_0} D(t,\tau) 
- \left( \frac{\mathcal{E}_1}{\mathcal{E}_0} \frac{2 U_p}{\omega_0} \right)^2   D^2(t,\tau)
\right] \, ,
\end{split}
\eeq
where we neglect one of the two quadratic terms, $i \mathcal{E}_1^2 \tau^3/24$, since it is dominated by the other one under the condition $U_p/ \omega_0 \gg 1$.

In the limit where the ionization potential of the generating system is much smaller than the ponderomotive energy of the freed electron, $I_p \ll U_p$, the integral~(\ref{x}) can be taken within the stationary point method as in~\cite{Lewenstein}. The stationary point $\tau=\tau_\mathrm{st}(t)$ corresponds to the zero value of the initial velocity of the electron $v(t-\tau)=p_\mathrm{st}(t,\tau)-A(t-\tau)=0$ for the quasi-classical action written in the form~\eqref{S_st_init}. Thus, the electron motion within this approximation is quasi-classical, and its features can be described within the simple-man picture~\cite{3-step_C,3-step_S}. Moreover, even if the condition $I_p \ll U_p$ is not valid, the main contribution to the integral~(\ref{x}) is given by the vicinity of the point $\tau=\tau_\mathrm{st}$, then the slowly-varying function $D(t,\tau)$ can be factored out from the integral as $D(t,\tau_\mathrm{st})$. In this case from Eqs.~(\ref{x}) and~(\ref{exp_S_st}) we can write the time-dependent dipole moment in the form
\beq{x012}
\begin{split}
x(t) = \ & x^{(0)}(t)+ 
i \frac{\mathcal{E}_1}{\mathcal{E}_0} \, \frac{2 U_p}{\omega_0} D(t,\tau_\mathrm{st}(t)) x^{(0)}(t) - \\
& \left [ \frac{\mathcal{E}_1}{\mathcal{E}_0} \, \frac{2 U_p}{\omega_0} D(t,\tau_\mathrm{st}(t))\right]^2 x^{(0)}(t) \, .
\end{split}
\eeq

The function $x^{(0)}(t)$ changes its sign every half-cycle of the laser field, and thus its spectrum consists of odd harmonics~\cite{Lewenstein}. From Eq.~(\ref{D}) one can see that function $D(t,\tau_\mathrm{st}(t))$ also changes its sign every laser half-cycle. Therefore, the second term in Eq.~(\ref{x012}) describes even harmonics, and the third one describes the correction of the odd harmonics due to the static field.
 
If now we assume that the field $\mathcal{E}_1$ is not a static one as in~\eqref{E}, but varies slowly with time,
\beq{E1quasistatic}
\mathcal{E}_1(t) = \mathcal{E}_1 \cos{(\omega_1 t)} \, , 
\eeq
then for
\beq{condition_freq}
\omega_1 \ll \omega_0 
\eeq
it can be considered as a quasi-static field. 
If we change the static field $\mathcal{E}_1$ in Eq.~\eqref{x012} with the quasi-static one~\eqref{E1quasistatic}, where $\cos{(\omega_1 t)}=[\exp(-i \omega_1 t)+\exp(i \omega_1 t)]/2$, this leads to the following alterations of the emitted spectrum. The linear term, $\propto \mathcal{E}_1$, causes the spitting of even harmonics into two satellites, shifted by $\pm \omega_1$ from the harmonic frequencies. The quadratic term $\propto \mathcal{E}_1^2$, meanwhile, results in the appearance of two satellites near the odd harmonics, shifted by $\pm 2 \omega_1$ from the harmonic frequencies, as well as a correction to the amplitude of the odd harmonics. 

These spectral changes can be related to the nonlinear high-order frequency mixing processes involving $q$ photons of the strong field $\mathcal{E}_0(t)$ and $m=1,2$ photons of the weak field $\mathcal{E}_1(t)$ with an odd total number of photons $q+m$. Using the notations from Ref.~\cite{Strelkov_np}, we describe these processes in terms of the induced susceptibilities $\kappa_q^{(m)}$ defined as the ratio of the spectral component of the atomic response to the corresponding power of the weak field: 
\beq{kappa_def}
\kappa_q^{(m)}\equiv x(\omega_2=q \omega_0 +m \omega_1)/ \mathcal{E}_1^{|m|} \, ,
\eeq 
where $x(\omega)$ is  the spectrum of the atomic response $x(t)$. 
Under conditions~(\ref{condition_amp}) and~(\ref{condition_freq}), $\kappa_q^{(m)}$ does not depend on either $\mathcal{E}_1$ or $\omega_1$ (see~\cite{Strelkov_np} for more details), but it does depend on $\mathcal{E}_0$ and $\omega_0$. Thus, single-photon processes in the weak low-frequency field with the sum and difference frequencies are designated as $m=1$ and $m=-1$, respectively; $m=2$ and $m=-2$ similarly denote two-photon processes.

The unperturbed harmonic response is denoted as~$\kappa_q^{(0)}$:
\begin{equation}
    \kappa_q^{(0)}\equiv 
    \begin{cases}
    x^{(0)}(\omega_2=q \omega_0) \, , & \text{for odd } q \\
    (\kappa_{q-1}^{(0)} + \kappa_{q+1}^{(0)})/2 \, , & \text{for even } q
    \end{cases}
     ,
\end{equation}
where $x^{(0)}(\omega)$ is the spectrum in the absence of the second field.

Finally, the quadratic correction to the HHG response in the weak low-frequency field is described with the susceptibility:
$$\kappa_q^{(0,2)} \equiv \left[ x(\omega_2=q \omega_0)-x^{(0)}(\omega_2=q \omega_0) \right] / \mathcal{E}_1^{2} \, .$$
This susceptibility describes a two-photon process, but without a frequency change.

From Eq.~(\ref{x012}) we derive the relative contributions of HFM processes in the two-color field in relation to the HHG process in the single-color field as
\begin{align}
\frac{\kappa_q^{(\pm 1)}}{\kappa_q^{(0)}} & = i  \frac{U_p}{\mathcal {E}_0 \omega_0} D (t_q,\tau_\mathrm{st}(t_q)) \, ,
\label{kappa1}
\\
\frac{\kappa_q^{(\pm 2)}}{\kappa_q^{(0)}} & = - \left[ \frac{U_p}{\mathcal {E}_0 \omega_0} D (t_q,\tau_\mathrm{st}(t_q)) \right]^2 \, ,
\label{kappa2}
\\
\frac{\kappa_q^{(0,2)}}{\kappa_q^{(0)}} & = - 2 \left[ \frac{U_p}{\mathcal {E}_0 \omega_0} D (t_q,\tau_\mathrm{st}(t_q)) \right]^2 \, .
\label{kappa02}
\end{align}
Here $t_q$ is the emission time of $q$\textsuperscript{th} harmonic within the simple-man picture, i.e., the return time of the electron which starts from the origin with zero initial velocity and returns with kinetic energy corresponding to the emission of XUV with photon energy close to $q \omega_0$. 
 
\begin{figure}
\raggedright
\includegraphics[width=0.95\linewidth]{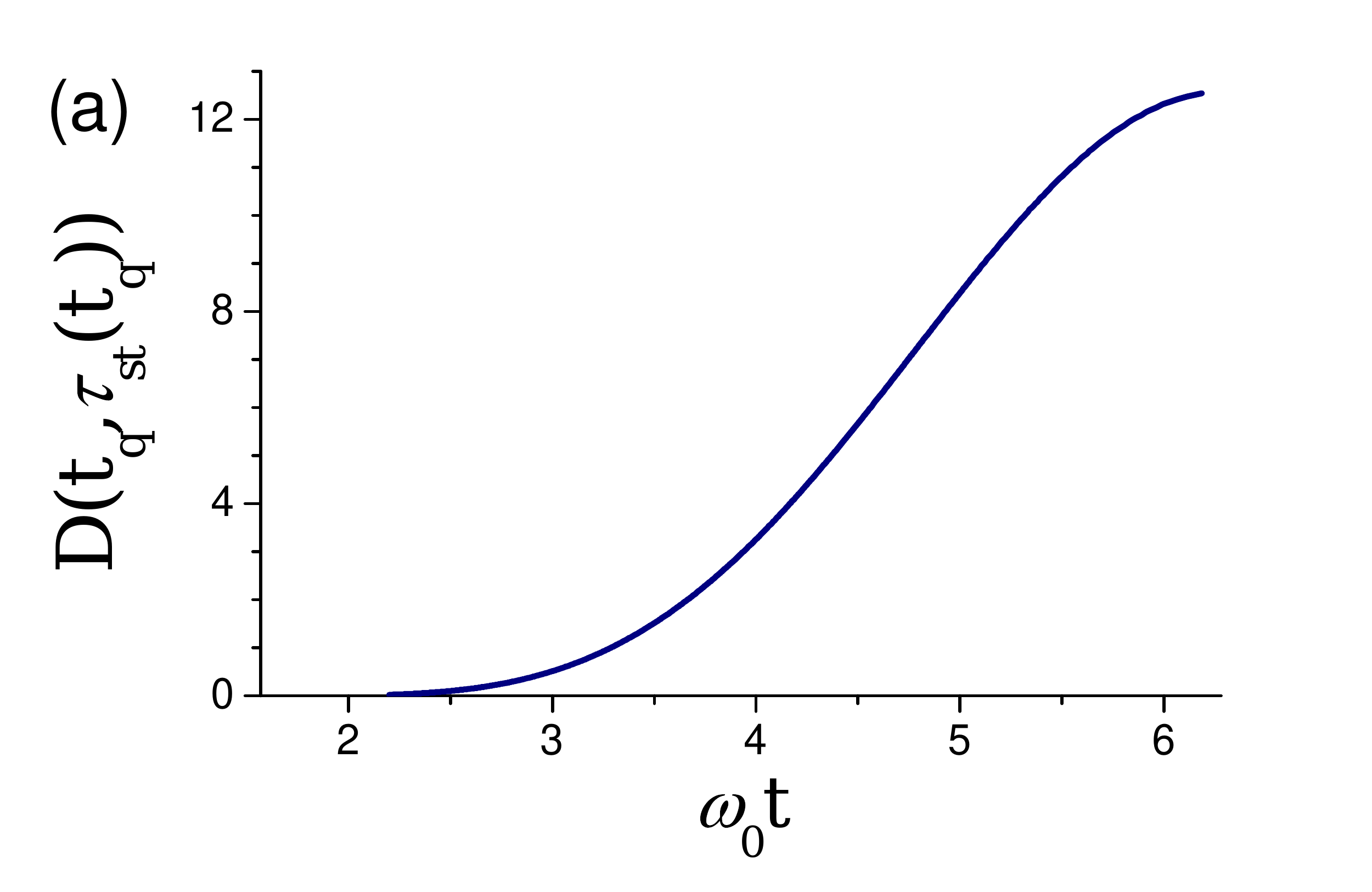}
\includegraphics[width=0.95\linewidth]{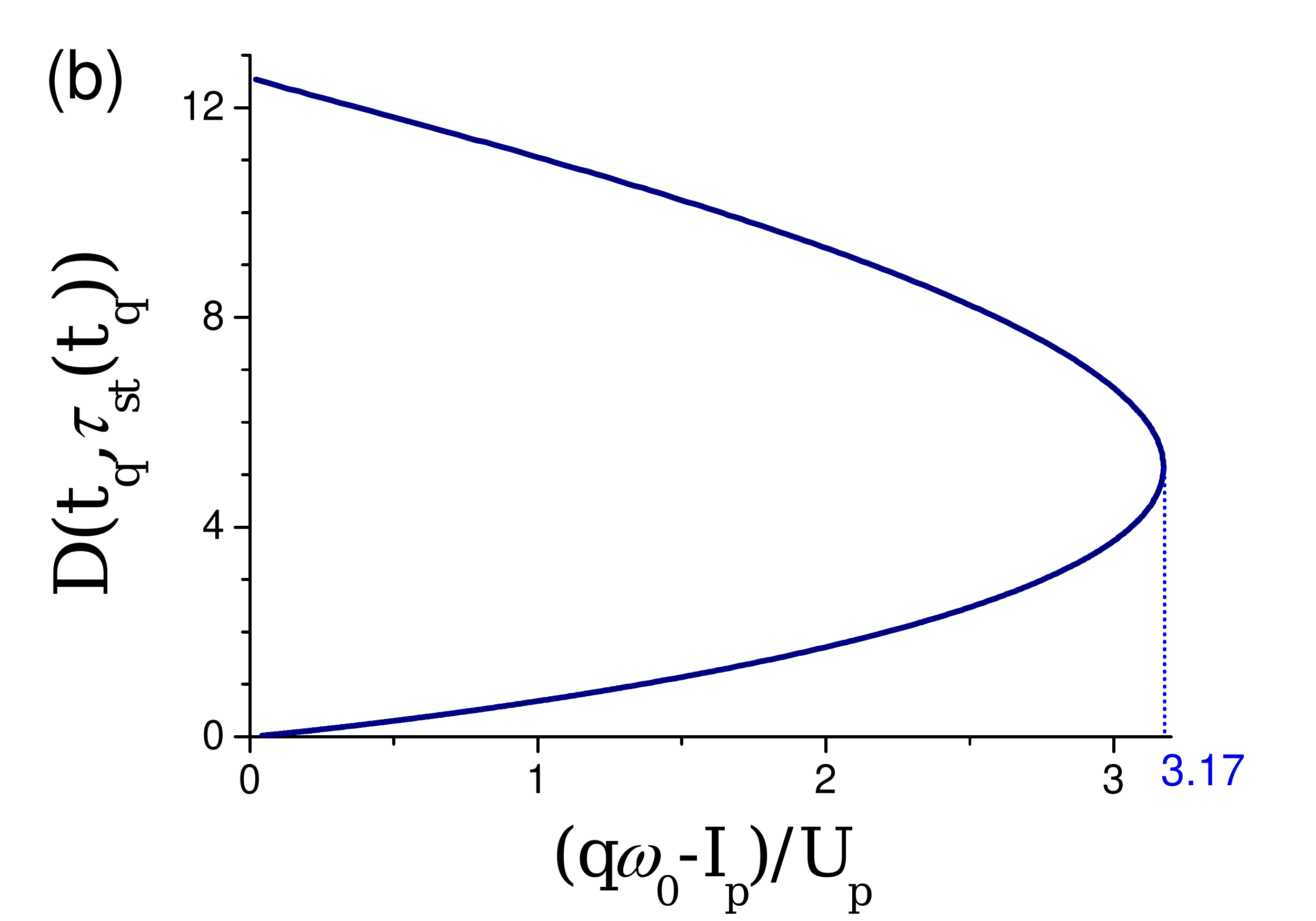}
\centering
\caption{The function $D(t,\tau)$ \eqref{D} for the electron quasi-classical trajectory started from origin with zero initial velocity as a function of~(a) the return time and~(b) the returning electron kinetic energy divided by the ponderomotive energy.}
\label{fig_D}
\end{figure}
 
One can see from Eqs.~\eqref{kappa1}~-~\eqref{kappa02} that contributions to the XUV spectrum involving one $\mathcal{E}_1(t)$ field photon are shifted in phase by $\pi/2$ with respect to the main contribution $\kappa_q^{(0)}$, while ones involving two photons are shifted by $\pi$. The latter can be understood as follows: the presence of the weak $\mathcal{E}_1(t)$ field should not change the total XUV emission efficiency by much; thus, the appearance of the new spectral components with some intensities should lead to a {\it decrease} of the ``main'' spectral component (i.e. the one with \mbox{$m=0$}). Note that a recent study of above-threshold ionization has shown similar behavior of phase shifts for contributions involving different number of photons~\cite{Bertolino2021}.

From Eqs.~\eqref{kappa1}~-~\eqref{kappa02} we also see that the dependence of the induced susceptibilities on the laser parameters is given by the factor $U_p /(\mathcal{E}_0 \omega_0)=\mathcal{E}_0/(4 \omega_0^3)$, while their dependence on the harmonic order is described by the function~$D(t_q,\tau_\mathrm{st}(t_q))$ calculated for the quasi-classical electron trajectory. Fig.~\ref{fig_D}(a) presents the value of $D(t_q,\tau_\mathrm{st}(t_q))$ as a function of the emission time and, in panel~(b), of the harmonic order. For the long electronic trajectory, the $D$ value is higher than for the short trajectory. This is natural because, the longer $\tau$ is, the stronger the influence of the \mbox{(quasi-)static} field on the electronic dynamics becomes. 

Below, we consider only the short trajectory contribution because this is the one usually observed experimentally. For this contribution the susceptibilities grow with the harmonic order, and this growth is more pronounced in the high-energy part of the plateau. Note that for the cut-off harmonics the assumption of a single quasi-classical trajectory is not valid (see, for instance, Ref.~\cite{Khokhlova} and references therein). So for the cut-off harmonics the behavior of the susceptibilities could differ from one shown in Fig.~\ref{fig_D}.
 
At a certain ``threshold'' field amplitude $\mathcal{E}_1^\mathrm{th}$ the contributions of the processes of different orders become equal, i.e., $|\kappa_q^{(m)} \mathcal{E}_1^\mathrm{th}| = |\kappa_q^{(m-1)}|$. From Eqs.~\eqref{kappa1}~-~\eqref{kappa02} we find this field as
\beq{E1th}
\mathcal{E}_1^\mathrm{th}=\mathcal{E}_0 \frac{\omega_0}{U_p D(t_q,\tau_\mathrm{st}(t_q))}=\frac{4 \omega_0^3}{\mathcal {E}_0 D(t_q,\tau_\mathrm{st}(t_q))} \, .
\eeq
Thus, if the laser field is intense (high $\mathcal{E}_0$) and has low enough frequency (low $\omega_0$), the field $\mathcal{E}_1$, even if very weak, still provides relatively intense HFM components. We also would like to stress the strong dependence of $\mathcal{E}_1^\mathrm{th}$ on the laser field frequency ($\propto \omega_0^3$).

For typical HHG conditions, namely, for laser-field intensity \SI{2e14}{W/cm^2} and \SI{800}{nm} wavelength, assuming $D=2$, we find the threshold intensity of the weak field as low as \SI{8e11}{W/cm^2}, which is $4 \times 10^{-3}$ of the laser intensity. Note that this level of mid-IR and THz field intensity is currently rather achievable.

\section{HFM in the frequency domain}
In this section we calculate spectral characteristics of the microscopic HFM response of a model argon atom in a two-color external field using numerically-integrated SFA and numerical TDSE solution (``SFA'' and``TDSE'' below, respectively), and compare these results with the analytical theory, which we call ``quasi-static SFA'' or ``qsSFA'', derived in the previous section for a qusi-static weak field. As above, here we consider the two-color field given by Eqs.~(\ref{E}) and~(\ref{E1quasistatic}) satisfying the conditions~(\ref{condition_amp}) and~(\ref{condition_freq}).

\subsection*{Methods}
To find the nonlinear atomic response via full SFA, we calculate the integral~(\ref{x}) numerically using the code~\cite{Emilio}. This approach allows one to separate different quantum path contributions for the plateau harmonics; here we study only the short quantum path contribution. Then we transform the time-dependent response to the spectral domain and use it to obtain the induced susceptibilities via Eq.~(\ref{kappa_def}). To obtain the nonlinear response from the TDSE, we solve the 3D TDSE via the numerical approach~\cite{Strelkov_2006} for a single-active electron (SAE) atomic potential~\cite{Strelkov_2005}, modeling an argon atom in the two-color field. 

For our calculations we use the following parameters of the external fields if not mentioned otherwise. The wavelength of the strong fundamental field is \SI{1200}{nm} and its intensity is \SI{2.4e14}{W/cm^2} for both SFA and TDSE calculations. For SFA, the frequency of the low-frequency field is $\omega_1=\omega_0/20$, the intensity is \SI{3e7}{W/cm^2}. We use a pulse consisting of 20 cycles, 3 cycles $\sin^2$ on-ramp, 14 cycles flat top and 3 cycles $\sin^2$ trailing edge. For TDSE, we use a higher frequency of the weak field $\omega_1=\omega_0/5$; this is done to have several oscillations of this field within time interval of the ground state depletion. We would like to note that we have checked that the numerical results are not sensitive to $\omega_1$ even at such (relatively high) values of this frequency. The intensity of the weak field used for TDSE is \SI{2.4e9}{W/cm^{2}} or $10^{-5}$ of the strong field intensity. Thus the weak-field intensity is higher than the one used in the SFA calculations~--- the higher intensity is chosen to minimize the numerical noise~--- but it is well below the threshold intensity
in the range \SI{e10}{}~- \SI{e11}{W/cm^2}, given by Eq.~(\ref{E1th}). The pulse duration is \SI{105}{fs}, the pulse consists of 10 cycles $\sin^2$ on-ramp, 20 cycles flat top and 10 cycles $\sin^2$ trailing edge.

\subsection{Intensity of HFM components}
\begin{figure}[]\centering
\includegraphics[width=0.98\linewidth]{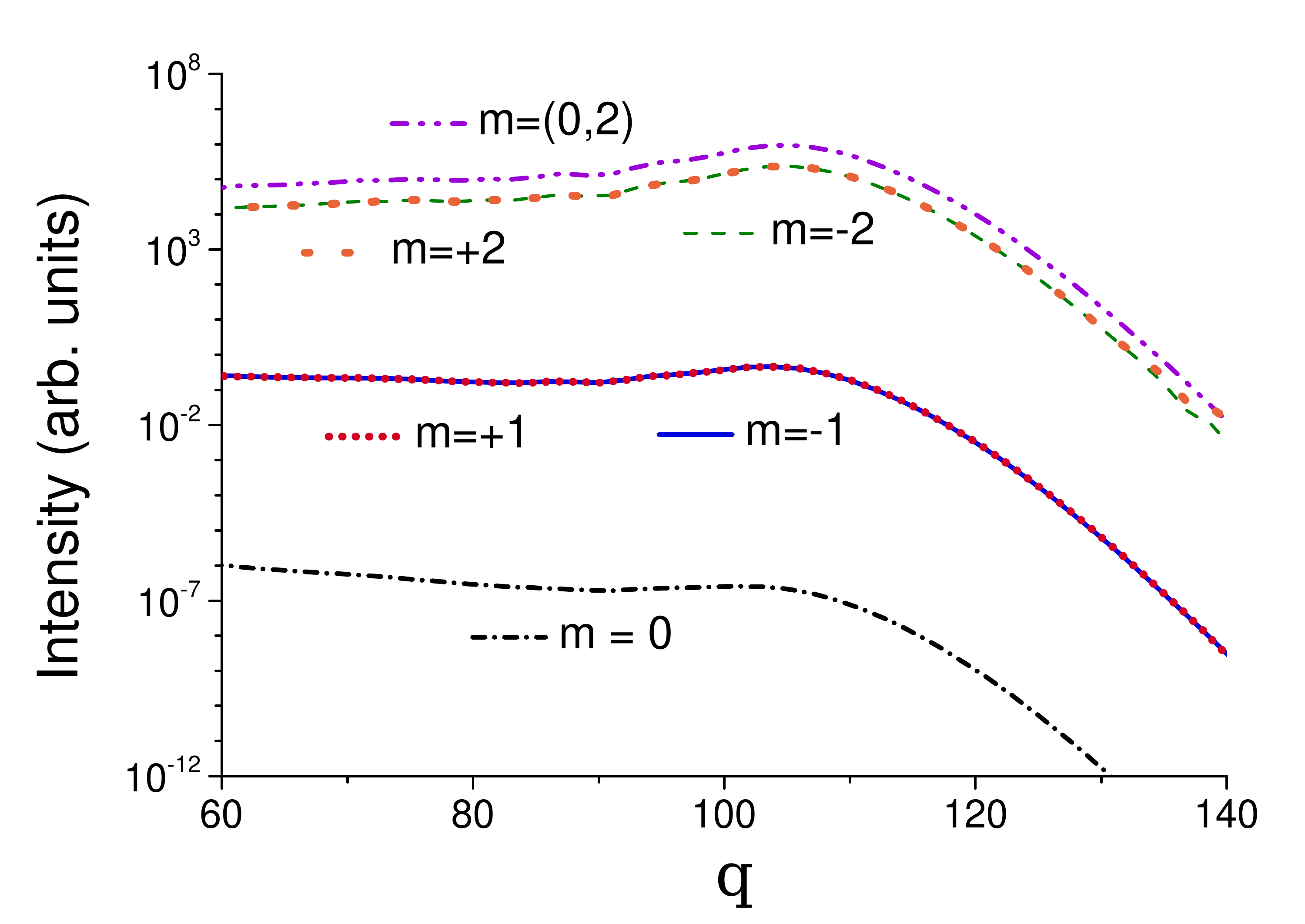} 
\caption{Relative intensity of susceptibilities $\kappa_q^{(m)}$ for HFM ($m \neq 0$) and HHG \mbox{($m=0$)} components as functions of the number of strong-field quanta $q$ for processes involving different numbers of the weak low-frequency field quanta $m$ calculated within SFA. See Methods for more details.
} 
\label{fig:kappa_vs_harm}
\end{figure}

We study the behavior of the spectral HFM response starting from the SFA results shown in \fig{fig:kappa_vs_harm} for the squared absolute values of the susceptibilities as a function of the number of strong-field photons $q$ for different numbers of the weak-field photons $m$.%
\footnote{%
We denote as $m=(0,2)$ the results related to $\kappa_q^{(0,2)}$.
}\ %
One can see that the susceptibilities are comparable for different $q$ lying in the plateau. Moreover, the susceptibilities for the same order $|m|$ are very close to each other $$ \kappa_q^{(m)} \approx \kappa_q^{(-m)} \, ,$$ which agrees with our analytical qsSFA results~(\ref{kappa1})-(\ref{kappa02}).

\begin{figure}\centering
\includegraphics[width=0.99\linewidth]{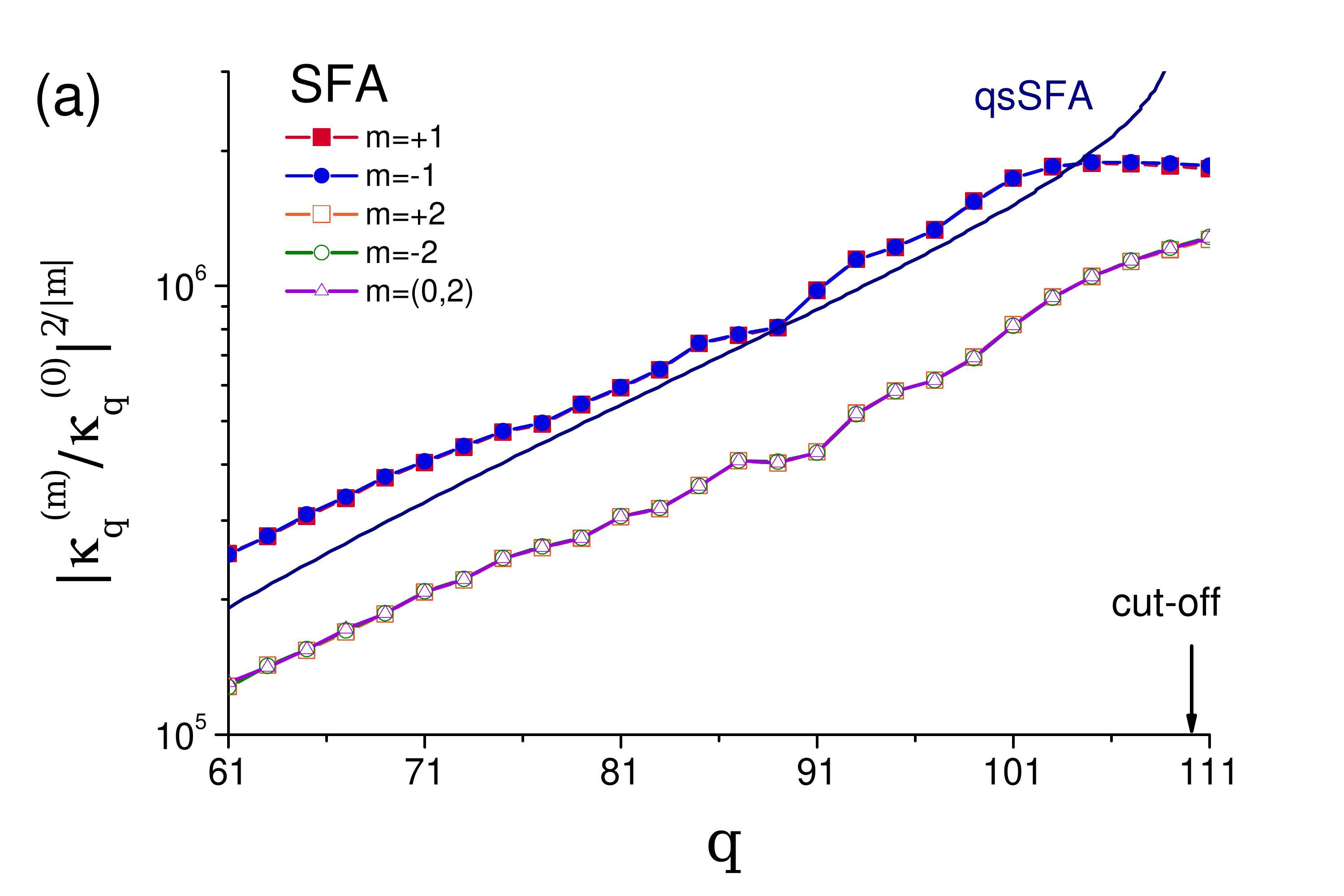} 
\includegraphics[width=0.99\linewidth]{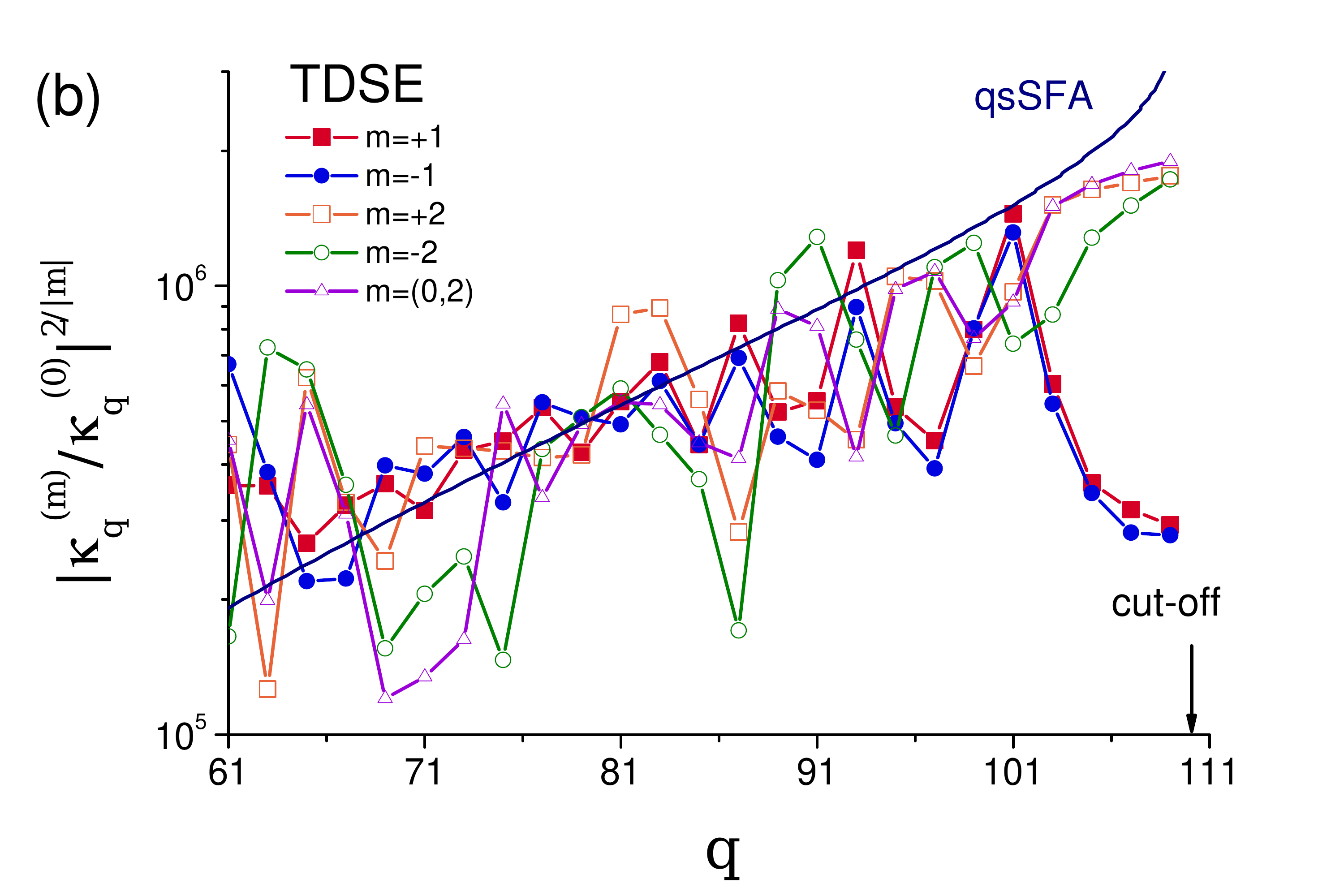} 
\caption{Ratios of the susceptibilities for $m \neq 0$ and $m=0$ as functions of $q$, calculated~(a) via SFA and~(b) via numerical TDSE solution. 
The analytical qsSFA result shown by navy solid line for all the ratios is $\left[ U_p/(\mathcal {E}_0 \omega_0) D (t_q,\tau_\mathrm{st}(t_q)) \right]^2$
.} 
\label{kappa_inten_ratio_vs_q}
\end{figure}

\fig{kappa_inten_ratio_vs_q} describes the ratio of the susceptibilities for $m \neq 0$ and $m=0$ obtained from SFA~(a) and TDSE~(b) calculations. For the first-order processes, $|m|=1$, we present the intensity ratio, while for the second order processes, $|m|=2$, we present the ratio of the absolute values; this corresponds to the analytical result for both cases [see Eqs.~(\ref{kappa1})~-~(\ref{kappa02})] being the same%
\footnote{%
For $m=(0,2)$ we present $\left| \frac{\kappa^{(0,2)}_q}{2 \kappa^{(0)}_q}\right|$.%
}%
: $\left[ \frac{U_p}{\mathcal {E}_0 \omega_0} D (t_q,\tau_\mathrm{st}(t_q)) \right]^2$. Panel~(a) demonstrates a good agreement of this analytical qsSFA result with the full SFA calculation, except for the cut-off region. The divergence in this region occurs due to inapplicability of the single-trajectory approach used for qsSFA. In turn, in panel~(b) one can see an overall agreement of the TDSE results for the ratio of the susceptibilities with the analytical qsSFA result for the plateau harmonics and again a weaker agreement for the cut-off ones.

\begin{figure}\centering
\includegraphics[width=0.99\linewidth]{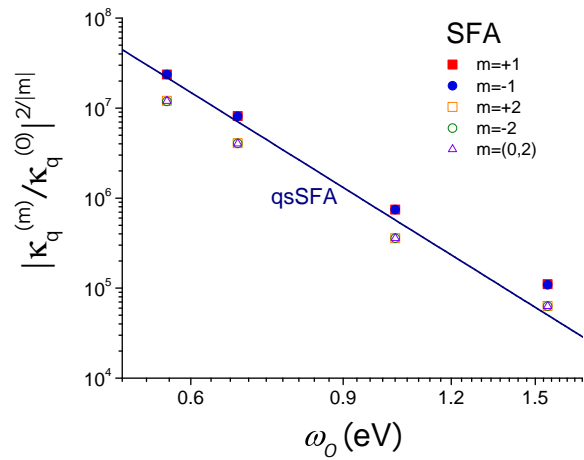}
\caption{Susceptibility ratio as functions of the fundamental frequency $\omega_0$ in log-log scale. These ratios are calculated via SFA for certain plateau harmonic, namely for $q$ order (depends on the fundamental frequency) corresponding to $D (t_q,\tau_\mathrm{st}(t_q))=2$.
} 
\label{sfa_kappa_inten_ratio_vs_omega_0}
\end{figure}

In~\fig{sfa_kappa_inten_ratio_vs_omega_0} we present the same ratios but as a function of the fundamental frequency calculated within SFA. These results show a reasonable agreement between SFA and qsSFA approaches for a wide range of frequencies.

\subsection{Phase of HFM components}
\begin{figure}[]\centering
\includegraphics[width=0.99\linewidth]{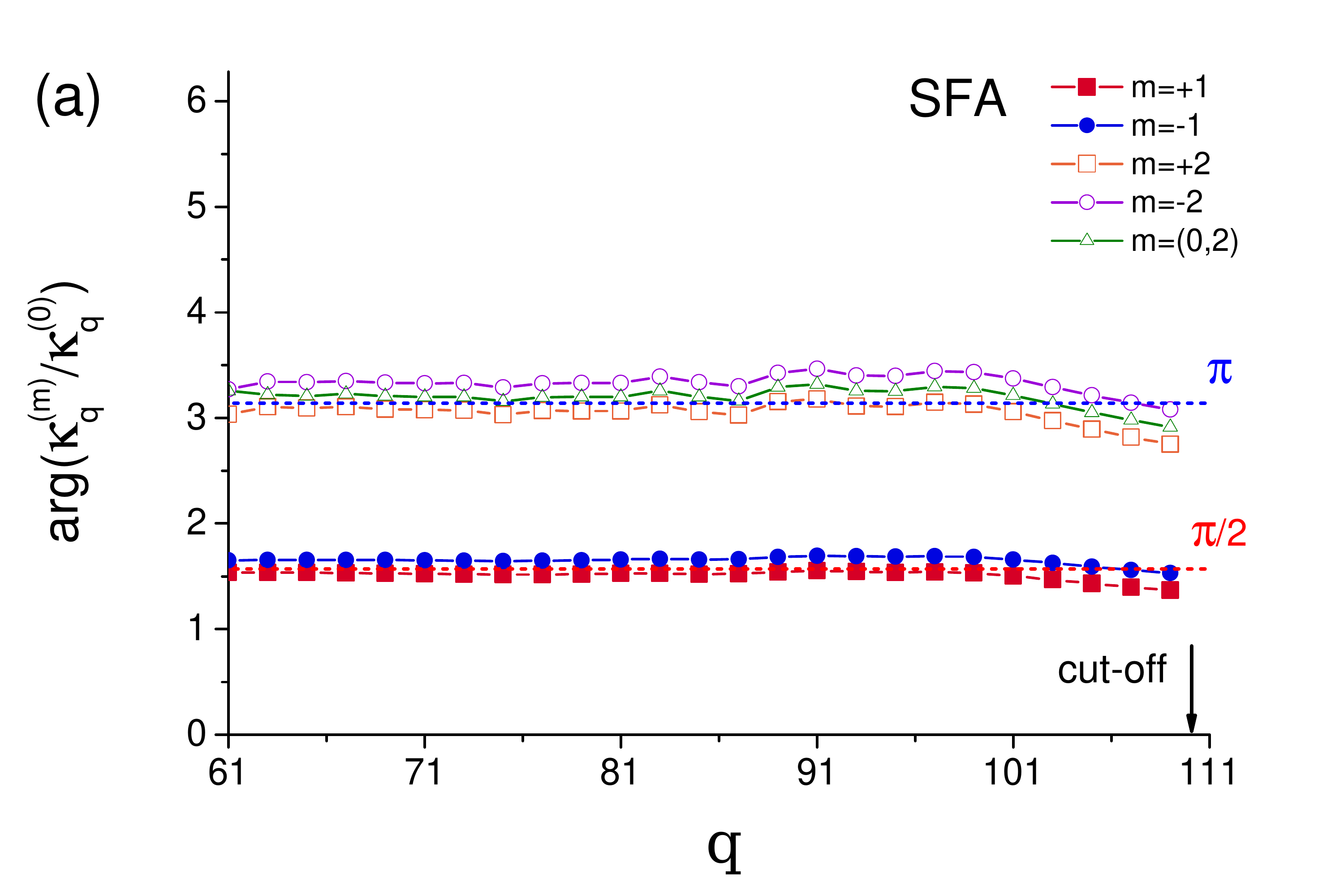}
\includegraphics[width=0.99\linewidth]{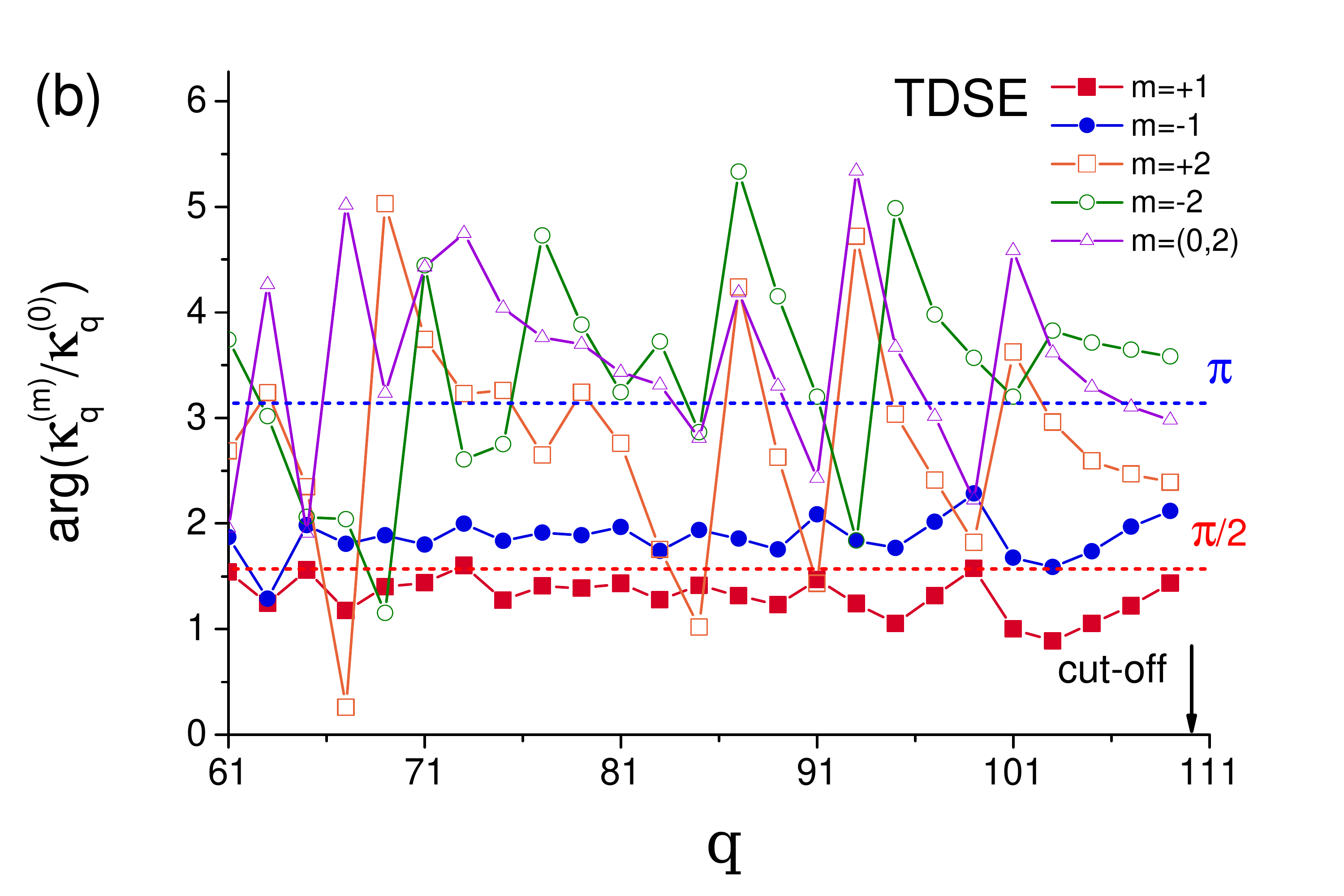}
\caption{Arguments of $\kappa_q^{(m)}/\kappa_q^{(0)}$ as functions of $q$ calculated via SFA~(a) and numerical TDSE solution~(b). The two-color field parameters are the same as in \fig{kappa_inten_ratio_vs_q}. Eq.~(\ref{kappa1}) predicts the argument of $\pi/2$ for $|m|=1$ and Eqs.~(\ref{kappa2}),~(\ref{kappa02})  predict the argument of $\pi$ for $|m|=2$}
\label{kappa_arguments}
\end{figure}

We analyze the behavior of the arguments of the susceptibility ratios as functions of $q$, see \fig{kappa_arguments}, calculated via SFA~(a) and numerical TDSE solution~(b). One can see a very good agreement between both SFA and TDSE, and qsSFA results for $|m|=1$ as well as between SFA and qsSFA results for $|m|=2$. The numerical TDSE results for $|m|=2$ are very noisy; however, the average result is still close to the analytical qsSFA prediction, except the cut-off region. The latter region performs a deviation discussed above.

\section{HFM in the time domain}
The difference between the spectral phases of the successive HHG components defines the emission time $t^\mathrm{e}_q=(\varphi_{q}-\varphi_{q-2})/(2 \omega_0)$ for attosecond pulses obtained using a group of harmonics close to the $q$\textsuperscript{th} one~\cite{Antoine1996,Platonenko1997,Mairesse2003,Varju2005}. The emission time for HHG components grows with increasing harmonic number~\cite{Ishikawa2010Feb} for plateau harmonics (the so-called harmonic ``attochirp'') and it is approximately constant for the cut-off ones. It has been also shown that the spectral region of these phase-matched harmonics expands with the fundamental frequency~\cite{Khokhlova}.

Assuming that the HFM components with certain $m$ can be selected experimentally (see the Discussion section), below we consider attosecond pulses obtained from a group of HFM components with given $m$ and different $q$. We generalize the concept of the emission time to HFM, where we define it through the phase difference between neighboring HFM components with the same $m$:
\beq{t_q^m}
t_q^{(m)}=\frac {\arg \left( \kappa_q^{(m)} \right)-\arg \left( \kappa_{q-2}^{(m)} \right) }{2 \omega_0} \, .
\eeq
 
Analytical qsSFA~(\ref{kappa1})~-~(\ref{kappa02}) as well as SFA and TDSE results in~\fig{kappa_arguments} show that the phase shift between $\kappa_q^{(m)}$ with different $m$ does not depend on $q$ in the plateau region. This means that the emission time for an attosecond pulse consisting of several HFM components with certain $m$ and different $q$ is the same as for the attosecond pulse obtained through the HHG process with corresponding $q$. Moreover, the attochirp for these pulses is the same, thus the duration of the attopulses is the same.

\begin{figure}[]
\centering
\includegraphics[width=1.0\linewidth]{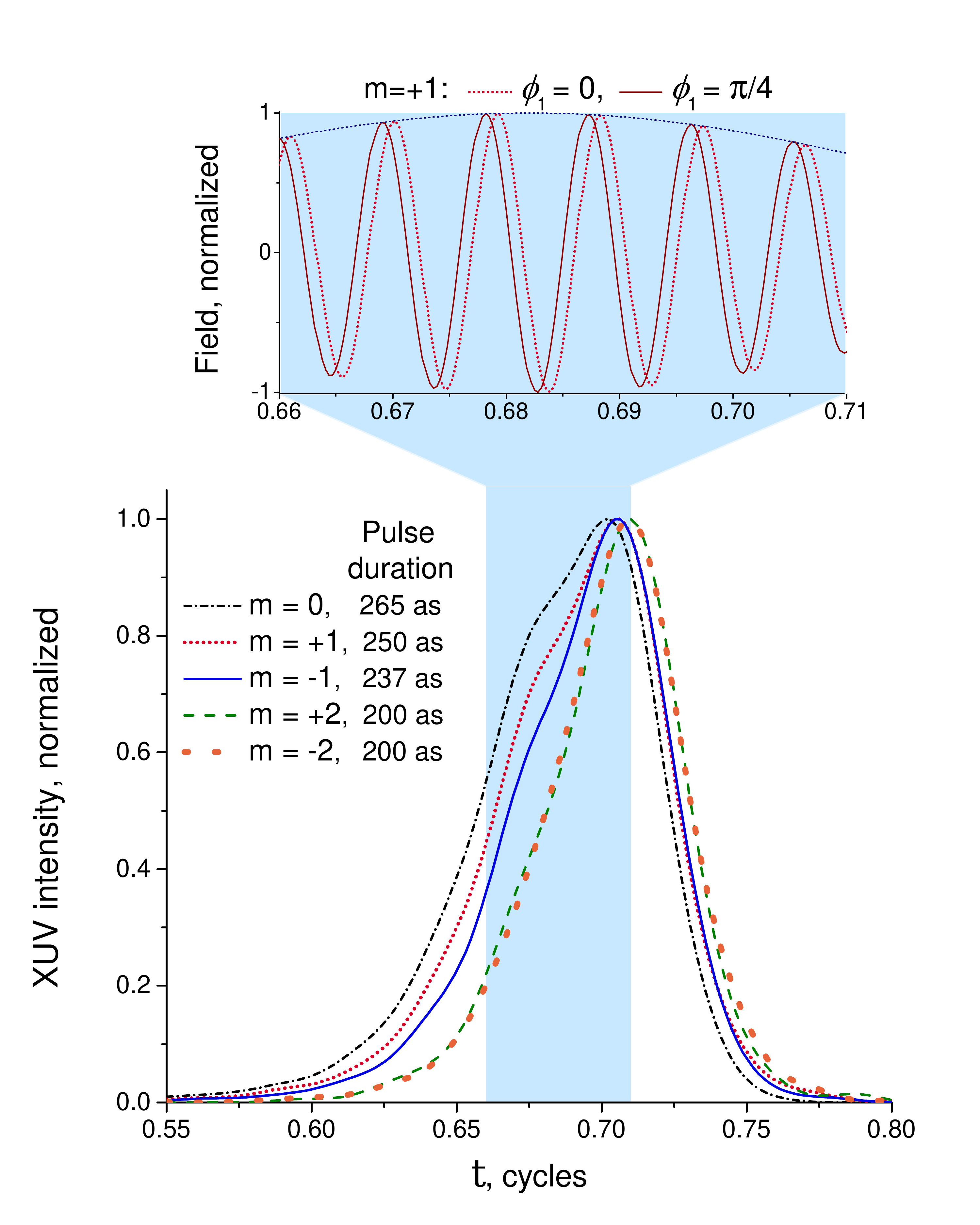} 
\caption{Normalized envelope of the attosecond pulses obtained via numerical TDSE solution using components with $q>100$ and different $m$; the attopulse durations are shown.
The inset shows the XUV fields of the attosecond pulses obtained using $m=+1$ for different phases of the weak field; the XUV field envelope is also shown.}
\label{attopulses}
\end{figure}

However, one can see in~\fig{kappa_arguments} that there is a regular difference from this behavior in the cut-off region. Calculating $t_q^{(m)}$ via Eq.~(\ref{t_q^m}) for HFM components near cut-off, we find that the spectral region where the HFM emission time does not depend on $q$ is broader for higher $m$ (for both SFA and TDSE results). Thus, the components in the cut-off region with higher $m$ can provide shorter attosecond pulses. \fig{attopulses} presenting the TDSE results (the SFA ones are similar) shows the normalized envelopes of the attosecond pulses obtained form the HFM components with the highest $q$. One can see that the attosecond pulse duration decreases with $|m|$. 

\begin{figure}[]
\centering
\includegraphics[width=0.95\linewidth]{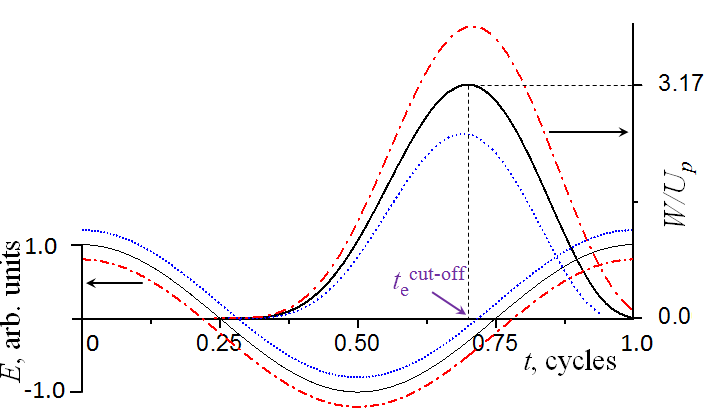}
\caption{Total field and kinetic energy of the returning electron calculated via the simple-man model as functions of time for zero quasi-static field (solid black lines) and for quasi-static field ``positive'' (dotted blue lines) and ``negative'' (red dash-dotted lines) to the laser field at the half-cycle when the electron is detached.}
\label{illustration}
\end{figure}

This feature of the cut-off region can be qualitatively explained considering the change of the HHG emission time caused by the quasi-static field within the simple-man approach. 
It is shown in \fig{illustration} that in the presence of the static field there are ``above-3.17 $U_p$'' harmonics generated due to the ionization at the ``negative'' half-cycle (i.e. when the laser field and the quasi-static field are opposite to each other), as was shown in Ref.~\cite{Taranukhin}. 
The emission time for these harmonics is close to the emission time for the cut-off harmonics generated in the absence of the static field, $t^\mathrm{e}_\text{cut-off}$. 
Moreover, for some range of ``below-$3.17 U_p$'' harmonics the emission time is also shifted towards $t^\mathrm{e}_\text{cut-off}$ because these harmonics correspond to the cut-off for the emission at the ``positive'' half-cycle. 
Thus, the correction of the emission time due to the quasi-static field moves this time towards $t^\mathrm{e}_\text{cut-off}$ for both ``above-3.17 $U_p$'' and ``below-$3.17 U_p$'' harmonics. 
As a result, the emission time of HFM components described by $\kappa_q^{(\pm2)}$ is close to $t^\mathrm{e}_\text{cut-off}$.

\section{CEP of attosecond pulses}
The CEP of the attosecond pulses obtained via HHG does not depend on the phase of the generating field, and depends on other properties of the generating pulse in a complex way~\cite{Guo2018, Sansone2006}. However, this phase is of key importance in some applications~\cite{Starace2007,Starace2013}. Here we show that the CEP of the attosecond pulses obtained via HFM can be easily controlled. It is curious to note that a similar CEP variation for femtosecond optical pulses obtained using a comb of frequencies slightly shifted from multiples of a repetition frequency is well-known in $f$-$2f$ interferometry~\cite{Paschotta}.

The field of the XUV attosecond pulse obtained using $m$\textsuperscript{th} order HFM components with complex amplitudes $\mathcal{E}_q^{(m)}$ is written as
\beq{atto_field}
\begin{split}
\sum_{q} \mathcal{E}_q^{(m)} &\exp(-i q \omega_0 t - i m \omega_1 t ) \equiv  \\
& \mathcal{E}_\mathrm{XUV}^{(m)}(t) \exp(- i \Omega t + i \varphi_\mathrm{CEP})  \, , 
\end{split}
\eeq  
where $\mathcal{E}_\mathrm{XUV}^{(m)}(t)$ is a slowly-varying periodic envelope, $\Omega$ is a carrier frequency and $\varphi_\mathrm{CEP}$ is a CEP. If the phases of the driving fields are changed by phase advances $\phi_0$~and~$\phi_1$, the phase of the HFM component changes by $q \phi_0 + m \phi_1$; here it is important to stress, that this is the case for arbitrary field amplitudes~\cite{Strelkov_np}, so the conclusions of this section are valid beyond the assumptions of weak amplitude and low frequency of the second field. The emission time of the attosecond pulse changes by $\delta t$, and its CEP changes by $\delta\varphi_\mathrm{CEP}$:
\beq{atto_field_phase_adv}
\begin{split}
\sum_q &\mathcal{E}_q^{(m)} \exp(-i q \omega_0 t +i q \phi_0 - i m \omega_1 t +i m \phi_1 ) \equiv       \\
&\mathcal{E}_\mathrm{XUV}^{(m)}(t+\delta t) \exp(- i \Omega (t+\delta t) + i \varphi_\mathrm{CEP}+i \delta \varphi_\mathrm{CEP}) \, .
\end{split}
\eeq  
Combining Eqs.~(\ref{atto_field}) and~(\ref{atto_field_phase_adv}), one finds
\beq{delta_t}
\delta t = - \phi_0 / \omega_0 
\eeq
and
\beq{delta_varphi_CEP}
\delta \varphi_\mathrm{CEP} = m(\phi_1- \phi_0 \frac{\omega_1}{\omega_0}) \, .
\eeq

This conclusion that phase advances of generating fields affect the CEP of attosecond pulse the generated via HFM is demonstrated by our numerical TDSE calculations. The inset in~\fig{attopulses} shows the fields of the attosecond pulses for $m=+1$ generated under $\phi_1=0$ and $\phi_1=\pi/4$. One can see that the CEP of the second pulse is shifted by $\pi/4$, in agreement with Eq.~(\ref{delta_varphi_CEP}). As a result, by tuning the phase of one of the generating fields, one can control the CEP of the attosecond pulse. 

For the attosecond pulses generated via HFM, the CEP varies for the successive pulses in the train, in contrast to the case of HHG. The variation of the CEP from one attosecond pulse to another $\Delta \varphi_\mathrm{CEP}$ can be found by writing the field~(\ref{atto_field}) at the time instant $t-T_0/2$ as
\beq{atto_field_shifted}
\begin{split}
\sum_q \mathcal{E}_q^{(m)}& \exp\left(-i q \omega_0 (t-T_0/2) - i m \omega_1 (t-T_0/2)\right) \equiv  \\
&\sum_q \mathcal{E}_q^{(m)} \exp(-i q \omega_0 t - i m \omega_1 t + i \Delta \varphi_\mathrm{CEP}) \, ,
\end{split}
\eeq  
where $T_0$ is the fundamental period. From the latter equation we have 
\beq{Delta_varphi_CEP}
\Delta \varphi_\mathrm{CEP} =
    \begin{cases}
        m \pi \frac{\omega_1}{\omega_0} \, , & \text{for odd } m \text{ (even } q\text{)} \\
        \pi + m \pi \frac{\omega_1}{\omega_0} \, , & \text{for even } m \text{ (odd } q\text{)}
    \end{cases}
    \, .
\eeq

One can notice that the change of the fundamental phase $\phi_0$ by $\pi$ corresponds to the change of the fundamental field direction to the opposite one, so the CEP of the attosecond pulse should change by $\pi$ (for even $m$) or by zero (for odd $m$). Let us show that this agrees with the equations above. The change of $\phi_0$ by $\pi$ leads to the CEP change according to Eq.~(\ref{delta_varphi_CEP}) by $\delta \varphi_\mathrm{CEP}=-m \pi \omega_1 / \omega_0$. According to Eq.~(\ref{delta_t}) this attosecond pulse is emitted at time $t+\delta t=t-T_0/2$, thus the attosecond pulse emitted at time $t$ is the next pulse in the train. Its CEP defined by Eq.~(\ref{Delta_varphi_CEP}) differs by $\Delta \varphi_\mathrm{CEP} = \pi + m \pi (\omega_1/\omega_0)$ or by $\Delta \varphi_\mathrm{CEP} = m \pi (\omega_1/\omega_0)$. Therefore, the total change of CEP $\delta \varphi_\mathrm{CEP}+\Delta \varphi_\mathrm{CEP}$ is equal to 0 or $\pi$.

\section{Discussion}
The HFM microscopic response for $m \ne 0$ is lower than the nonlinear microscopic response for HHG, at least within the fields $\mathcal{E}_1<\mathcal{E}_1^\mathrm{th}$ considered here. However, the macroscopic response for HFM with $m<0$ can be much higher due to significantly better phase matching~\cite{KhokhlovaStrelkov}. This takes place for a certain frequency ratio of the generating fields and a certain $m$, defined by this ratio. Thus, the phase matching should provide high generation efficiency in conjunction with selection of the HFM components with desired $m$. The investigation of the macroscopic HFM properties is a natural outlook of the present study.

There is another approach to separate the HFM components with different $m$ based on the use of non-coaxial generating beams~\cite{Worner_non-collinear, Chappuis}. Within this approach the HFM components with negative $m$ (effectively generated due to better phase matching) are emitted in a direction different from the directions of the generating beams. This makes detection and utilization of this radiation more convenient. Finally, co-axial generating beams with different focusing properties can be used to obtain focused HFM beams with certain $m$. 

 HFM paves a way to the generation of a single attosecond pulse with controllable CEP. Namely, if the laser field provides some gating for attosecond pulse generation (such as ellipticity gating~\cite{Sola} or attosecond lighthouse~\cite{Vincenti2012}), this allows for isolated attosecond pulse generation. The phase variation of the weak generating field would not affect the gating properties, but it would provide CEP control for the generated single attosecond pulse. 
\section{Conclusion}
We investigate theoretically the single-atom properties of the HFM process for the case of a strong laser field combined with a weaker low-frequency one. 
Using SFA theory we consider the latter field as a quasi-static one, and assume that the main role of this field is to produce a correction of the action accumulated by the electron during its free motion. 
Within this assumption we show that the amplitudes of HFM spectral components generated by $q$ fundamental photons and $m$ low-frequency photons ($|m| \le 2$ are considered) can be written as a product of the $q$\textsuperscript{th} high-harmonic amplitude, the $|m|$\textsuperscript{th} power of the weak field amplitude, and a multiplier which increases with $\tau$ (the time of the electronic free motion) and rapidly decreases with the fundamental frequency, see Eqs.~(\ref{kappa_def})~-~(\ref{kappa02}). 
We show that the HFM components are shifted in phase by $|m| \pi/2$ with respect to high harmonics. 
For $q$ lying in the plateau region these analytical results agree with numerically-integrated SFA, as well as with numerical TDSE simulations, while for the cut-off region there is a regular deviation. 

This deviation describes the spectral region of attochirp-free HFM components which is broader than the one for HHG. 
We discuss the origin of this feature of the cut-off HFM components and demonstrate that it leads to shorter durations of attosecond pulses obtained via HFM, and that the duration decreases with an increase of $|m|$.
Moreover, we show that the CEP of the attosecond pulses obtained via HFM can be easily controlled by tuning the phases of the generating fields, while such control is impossible for the pulses obtained via HHG. 
The equations describing the attosecond pulse CEP are applicable even beyond the assumption of weakness and low frequency of the second field. 
Finally, we would like to stress that due to perspective of phase-matched generation for long propagation distances, HFM can substantially improve the efficiency of the attosecond pulse sources.

\section*{Acknowledgments}
We acknowledge funding from ``Basis'' Foundation for the Advancement of Theoretical Physics and Mathematics. 
The TDSE calculations and the code development were funded by RSF (grant No 22-22-00242). 
M.K. acknowledges funding from the Alexander von Humboldt Foundation.

\bibliography{lit}

\end{document}